\documentclass[aps,twocolumn,superscriptaddress,prb]{revtex4-2}
\usepackage{graphicx,amsmath}
\usepackage{amssymb}
\usepackage{enumitem}
\usepackage{bbm}
\usepackage[usenames,dvipsnames]{color}
\usepackage{xcolor}
\usepackage{color}
\usepackage[normalem]{ulem}

\newcommand{\be}{\begin{equation}}
\newcommand{\ee}{\end{equation}}

\newcommand{\bs}{\boldsymbol}

\begin{document}

\title{Floquet Engineering of Lie Algebraic Quantum Systems}

\author{Jayendra N. Bandyopadhyay}
\email{jnbandyo@gmail.com}
\affiliation{Center for Theoretical Physics of Complex Systems, Institute for Basic Science (IBS), Daejeon 34126, Republic of Korea}
\affiliation{Department of Physics, Birla Institute of Technology and Science, Pilani 333031, India}
\author{Juzar Thingna}
\email{jythingna@ibs.re.kr}
\affiliation{Center for Theoretical Physics of Complex Systems, Institute for Basic Science (IBS), Daejeon 34126, Republic of Korea}
\affiliation{Basic Science Program, University of Science and Technology (UST), Daejeon 34113, Republic of Korea}


\begin{abstract}

We propose a `Floquet engineering' formalism to systematically design a periodic driving protocol in order to stroboscopically realize the desired system starting from a given static Hamiltonian. The formalism is applicable to interacting and non-interacting quantum systems which have an underlying closed Lie-algebraic structure. Unlike previous attempts at Floquet engineering, our method produces the desired Floquet Hamiltonian at any driving frequency and is not restricted to the fast or slow driving regimes. The approach is based on Wei-Norman ansatz, which was originally proposed to construct a time-evolution operator for any arbitrary driving. Here, we apply this ansatz to the micro-motion dynamics, defined within one period of the driving, and engineer the functional form and operators of the driving protocol by fixing the gauge of the micro-motion. To illustrate our idea, we use a two-band system or the systems consisting of two sub-lattices as a testbed. Particularly, we focus on engineering the cross-stitched lattice model that has been a paradigmatic flat-band model.

\end{abstract}

\maketitle

\noindent \emph{Introduction}.-- Floquet formalism~\cite{Floquet1883} has been instrumental to study the dynamic evolution of a system subjected to a periodic driving. The dynamics is decomposed into two parts a time-periodic part describing the micro-motion of the system within a period, and an effective stroboscopic part governed by a static `Floquet Hamiltonian'. The problem of reverse engineering the driving protocol in order to obtain a desired Floquet Hamiltonian stroboscopically from a given {\it simple} static Hamiltonian is known as \emph{Floquet engineering}. It has garnered a lot of attention over the past several years and has been applied in different experimental paradigms \cite{Floquet-expt01,Floquet-expt02,Floquet-expt03,Floquet-expt04,Floquet-expt05,Floquet-expt05}. Floquet engineered solid-state materials have been discussed extensively to develop ``quantum matter on demand" by controlling post-semiconductor materials \cite{Floquet-Matt03,Floquet-Matt04,Floquet-Matt05} and several \emph{exotic} properties like unconventional superconductivity \cite{Floquet-Matt01,Floquet-Matt02}, topologically nontrivial band structures \cite{Rechtsman2013}, etc. have been realized. Moreover, the effect of periodic driving has been studied on a variety of timely solid-state systems such as Luttinger liquid \cite{Luttinger}, superconducting circuit \cite{SuperCircuit}, bilayer graphene \cite{BLG}, and strongly correlated electrons (Mott materials) \cite{Mentink2015}.

Most of these studies investigated the effect of periodic driving, using a square-wave or sinusoidal protocol, on a given system in either the high~\cite{Goldman-Dalibard,Eckardt,ShiraiNJP16,BW-Oka,EckardtRMP} or low~\cite{LowFreq01,LowFreq02} driving frequency regime. However, a systematic theory of designing a driving protocol such that the desired Floquet Hamiltonian obtained exactly at \emph{any} driving frequency is still missing in the literature. 

In this Letter, we propose to bridge this gap by formulating a theory of Floquet engineering for a class of systems whose Hamiltonians have any closed Lie-algebraic structure. Our formalism is based on the Wei-Norman ansatz, which was originally proposed to obtain the dynamics for any time-dependent system \cite{WeiNorman-PAMS,WeiNorman-JMP}. Since the form of the long-time evolution part is already known from Floquet theory, we massage the Wei-Norman ansatz to the micro-motion part of the dynamics. 

The requirement of a Lie-algebraic structure is not a drawback, but Hamiltonians of several important classes of solid state systems obey this structure. Hamiltonians for non-interacting two-band systems in any dimension follow the $SU(2)$ algebra \cite{girvin}. Even interacting one- (two-) dimensional models of unconventional chiral $p$-wave superconductors at the mean-field level \cite{SI,Kitaev2001,bernevig-book,de-Gennes} or Weyl semi-metals \cite{Weyl-SemiMetal} belong to the $SU(2)$ class. Three-band systems like the Kagome and Lieb lattice obey $SU(3)$ algebra \cite{SI}. Furthermore, high-$T_c$ superconductors that can be modeled by a half-filled Hubbard Hamiltonian have an underlying $SO(4)$ symmetry. The Bethe ansatz solution of these models show that its elementary excitation can be separated into two fundamental excitations: spinons and holons/antiholons, which reflects that the original $SO(4)$ symmetry can be separated into $SU(2) \otimes SU(2)$ symmetry \cite{SU2XSU2-01,SU2XSU2-02}. Another direct example of the importance of the Lie-algebraic structure is a strongly interacting system described by a Tomonaga-Luttinger liquid that obeys $SU(1,1)$ algebraic structure and has been studied under periodic driving \cite{Floquet-Luttinger}. 

Our formalism can not only be used to control the dynamics of these systems in presence of driving but could also help design the \emph{full} driving protocol to realize these complex Hamiltonians stroboscopically using simple static Hamiltonians~\cite{SI}. We first outline our general formalism and then illustrate the main idea by designing a driving protocol to realize the {\it cross-stitched} lattice, an interesting two-band system with one band dispersive and another flat, from a static site Hamiltonian~\cite{Flatband01,Flatband02}.

\noindent\emph{Formalism}.-- The Hamiltonian of a generic periodically driven quantum system reads  
\be
H(t) = H_0 + V(t), ~~V(t+T) = V(t),
\ee
where $H_0$ and $V(t)$ are the static Hamiltonian and the driving potential with time-period $T$, respectively. The corresponding time-dependent Schr\"odinger equation (TDSE) is
\be
i \frac{dU(t)}{dt} = H(t)\, U(t),~~~~(\hbar=1).
\label{eq:TDSE}
\ee
The operator $U(t)$ is the unitary time-evolution operator. According to the Floquet theorem, the solution of the TDSE can always be expressed as
\be
U(t) = P(t)\, e^{-i H^{\rm eff} t},
\label{eq:TDSE-Schr}
\ee
where the {\it micro-motion operator} $P(t+T) = P(t)$ describes the dynamics of the system within one period $[t, t+T]$ and  $H^{\rm eff}$ is a static Hamiltonian that governs the long-time dynamics of the system. The initial condition $U(0) = \mathbbm{1}$ imposes the condition $P(0) = \mathbbm{1}$, whereas the time-periodicity gives $P(nT) = \mathbbm{1}$ for every $n \in \mathbbm{Z}^+$ (positive integers). Consequently, we have $U(nT) = e^{-i H^{\rm eff} nT} = [e^{-i H^{\rm eff}T}]^n = [U(T)]^n$. If the dynamics of the system is observed stroboscopically at $t=nT$, then it is governed by the effective static Hamiltonian $H^{\rm eff}$.

Obtaining the analytic quantum evolution for any Hamiltonian is highly nontrivial and hence we restrict ourselves to those Hamiltonians whose operators form a closed Lie-algebra, i.e.,
\begin{eqnarray}
H_0 &=& h_0 \mathbbm{1} + \sum_{\alpha=1}^N h_\alpha A_\alpha = h_0 \mathbbm{1} + \boldsymbol{h} \bs{\cdot} \boldsymbol{A},\\
V(t) &=& f_0(t) \mathbbm{1} + \sum_{\beta=1}^N f_\beta(t) A_\beta = f_0(t) \mathbbm{1} + \boldsymbol{f}(t) \bs{\cdot} \boldsymbol{A},
\end{eqnarray}
where $\bs{\cdot}$ denotes the standard scalar product. Above, $h_0$ and $\bs{h}$ (column vector with elements $h_{\alpha}$ and dimension $N$) are time-independent parameters, whereas $f_0(t)$ and $\bs{f}(t)$ are time-dependent functions due to the external field. The column vector of the linear operators $\bs{A}=\{A_{\alpha}\}$ forms a finite $N$-dimensional {\it simple} Lie-algebra $\mathcal{L}_N$, which satisfies
\be
[A_\alpha, A_\beta] = \sum_{\gamma=1}^N \Lambda_{\alpha\beta}^{\gamma} A_\gamma,
\ee
where $\Lambda$'s are the structure constants of the algebra $\mathcal{L}_N$. From the Floquet engineering perspective, the underlying Lie-algebraic structure will be exploited to design a driving scheme $\{f_0(t),\bs{f}(t)\}$ to achieve a desired effective Hamiltonian $H^{\rm eff}$ for a given initial static Hamiltonian $H_0$.

The Wei-Norman ansatz \cite{WeiNorman-JMP,WeiNorman-PAMS}, i.e., expressing the full evolution operator as a product of exponentials, has been successfully applied to solve the TDSE for a driven quantum system. In our case, since we are particularly interested in Floquet engineering wherein $H^{\rm eff}$ is known, the natural choice is to apply the ansatz to the micro-motion operator,
\be
P(t) = e^{-i m_0(t)}\, \left(\prod_{\alpha=1}^N e^{-i m_\alpha(t) A_\alpha}\right).
\label{eq:ansatz}
\ee
The initial condition and the time-periodic property of $P(t)$ imposes following conditions: $m_0(nT) = 2\nu n\pi$ and $e^{-i m_\alpha(nT) A_\alpha} = \mathbbm{1}$ for all $\alpha = 1, \cdots, N$, $n=0,1,2,\cdots$, and any integer $\nu$. Besides, we have a gauge freedom to choose any time-dependent functional form of $m_\alpha(t)$. Using the above form of $P(t)$, if we substitute $U(t)$ in the TDSE [Eq. \eqref{eq:TDSE}], we get the relations between the driving protocols $\{f_0(t), \bs{f}(t)\}$ and the functions $\{m_0(t), \bs{m}(t)\}$ as
\be
\begin{split}  
&[h_0 + f_0(t)] + [\bs{h} + \bs{f}(t)] \bs{\cdot} \bs{A} \\ &~~~~~ = \dot{m}_0(t) + \bs{\zeta}({\bs m}, \dot{\bs{m}}) \bs{\cdot} \bs{A}  + P(t)\, H^{\rm eff}\, P^\dagger(t).
\end{split}
\label{eq:f-m01}
\ee
Here, the components of the column vector $\bs{\zeta}$ are linear functions of $\dot{\bs{m}}(t)=\{dm_{\alpha}(t)/dt\}$ and nonlinear functions of $\bs{m}(t)$. Therefore, we can always express $\bs{\zeta}({\bs m},\dot{\bs{m}}) = \bs{{\mathcal M}}_1(t) \bs{\cdot}\dot{\bs{m}}$, where $\bs{{\mathcal M}}_1(t)$ is a $N \times N$ matrix whose elements are nonlinear functions of $\bs{m}$. This nonlinearity is decided by the underlying Lie-algebra. Consider the general form $H^{\rm eff} = h^{\rm eff}_0 \mathbbm{1} + \bs{h}^{\rm eff} \bs{\cdot} \bs{A}$, the last term on the right hand side of Eq. \eqref{eq:f-m01} can also be represented in terms of the operators $\bs{A}$ as
\be
P(t)\, H^{\rm eff}\, P^\dagger(t) = h^{\rm eff}_0 \mathbbm{1} +\bs{\xi}(\bs{m}, \bs{h}^{\rm eff}) \bs{\cdot} \bs{A}.
\label{eq:f-m02}
\ee
The vector $\bs{\xi}(\bs{m}, \bs{h}^{\rm eff})$ is a linear function of $\bs{h}^{\rm eff}$, but a nonlinear function of $\bs{m}(t)$, i.e., $\bs{\xi}(\bs{m}, \bs{h}^{\rm eff}) = \bs{{\mathcal M}}_2(t) \bs{\cdot}\bs{h}^{\rm eff}$ where $\bs{{\mathcal M}}_2(t)$, similar to $\bs{{\mathcal M}}_1(t)$, is a matrix whose elements are nonlinear functions of $\bs{m}$. Using Eqs.~\eqref{eq:f-m01} and~\eqref{eq:f-m02} and equating the coefficients of the operators, we get
\begin{eqnarray}
\label{eq:equate_coeff}
h_0 + f_0(t) &=& \dot{m}_0(t) + h^{\rm eff}_0,\\
\bs{h} + \bs{f}(t) &=& \bs{\mathcal{M}}_1(t) \bs{\cdot}\dot{\bs{m}}(t) + \bs{\mathcal{M}}_2(t) \bs{\cdot}\bs{h}^{\rm eff}, \nonumber
\end{eqnarray}  
where $\bs{\mathcal{M}}_1(nT) = \bs{\mathcal{M}}_2(nT) = \mathbbm{1}$ for $n = 0, 1, 2, \cdots$. The gauge freedom in the micro-motion operator makes $\bs{\mathcal{M}}_1(t)$ and $\bs{\mathcal{M}}_2(t)$ non-unique, but it can be fixed at any arbitrary time $t \neq nT$ by choosing an appropriate gauge. According to Wei-Norman, if $\mathcal{L}_N$ is not a solvable algebra, the transformation matrices $\bs{\mathcal{M}}_1(t)$ and $\bs{\mathcal{M}}_2(t)$ could be ill-defined for an arbitrary representation. Therefore, unless we find a representation which is globally well-defined, we cannot apply the Wei-Norman ansatz to design the driving protocol.

Our Lie-algebraic Floquet engineering protocol can be applied to any system having an underlying finite dimensional closed algebra. We now apply this formalism to an arbitrary two-bands system that naturally follows the $SU(2)$ algebra. Here we are particularly focusing on this algebra because a large class of interacting and noninteracting systems obey this symmetry \cite{Kitaev2001,SI,bernevig-book,de-Gennes,girvin,Volovik1999,Read-Green,Ivanov}. In principle, this formalism can also be applied to multi-bands systems, but the complexity of the problem increases with the number of bands (see supplementary~\cite{SI} for the three-band case).      

\noindent\emph{Two-bands systems}.-- In the momentum space ($\bs{k}$-space), the Hamiltonian of the periodically driven $2$-band systems can be written in terms of the Nambu spinors $\Psi_{\bs{k}} = \left(a_{\bs{k}},b_{\bs{k}}\right)^{\mathrm{T}}$ as
\be 
H(t) = \sum_{\bs{k}} \Psi_{\bs{k}}^\dagger H_{\bs{k}}(t) \Psi_{\bs{k}},
\label{eq:su2-ham}
\ee
where $H_{\bs{k}}(t) = H_{\bs{k}0} + V_{\bs{k}}(t)$ and $V_{\bs{k}}(t+T) = V_{\bs{k}}(t)$. The components of the Nambu spinors $a_{\bs{k}} (a_{\bs{k}}^\dagger)$ and $b_{\bs{k}} (b_{\bs{k}}^\dagger)$ are respectively representing the annihilation (creation) operators corresponding to the valence and conduction band. The time-independent part $H_{\bs{k}0}$ and the time-periodic $V_{\bs{k}}(t)$ can be expressed in general as,
\begin{eqnarray}
H_{\bs{k}0} &=& h_{\bs{k}0} \mathbbm{1} + \bs{h}_{\bs{k}} \bs{\cdot} \boldsymbol{S}\nonumber \\
V_{\bs{k}}(t) &=& f_{\bs{k}0}(t) \mathbbm{1} + \bs{f}_{\bs{k}}(t) \bs{\cdot} \bs{S}.
\label{eq:Ham-comp}
\end{eqnarray}
The operators $2\bs{S}= \bs{\sigma}$ follow $SU(2)$ algebra, where the components of $\bs{\sigma}$ are the Pauli matrices. This finite dimensional algebra facilitates the application of the Wei-Norman formalism to study the dynamics of two-band systems. 

The unsolvable $SU(2)$ algebra has two well-known representations: $XYZ$ representation with $\boldsymbol{S}_{XYZ} = (S_x, S_y, S_z)^{\mathrm{T}}$ and $\pm Z$ representation with $\boldsymbol{S}_{\pm Z} = (S_+,S_-,S_z)^\mathrm{T}$, where $S_\pm = (S_x \pm i S_y)$. For an arbitrary choice of $\bs{A}$, e.g., $\bs{A} = \bs{S}_{XYZ}$, it is not guaranteed that the time-dependent functions $m_\alpha (t)$ appearing in the micro-motion operator, Eq. \eqref{eq:ansatz}, are smooth continuous functions for all time $t$~\cite{WeiNorman-PAMS,WeiNorman-JMP}. However, following Ref. [\onlinecite{ARPRau-PRL}], we later show that the $(\pm Z)$ representation gives a globally well-defined $\bs{\mathcal M}_1(t)$ matrix. Therefore, for the two-bands case, Wei-Norman ansatz along with a proper choice of representation $\bs{S} \equiv \bs{S}_{\pm Z}$ can be applied to design a driving scheme, with arbitrary driving frequency, to achieve the desired effective Hamiltonian from a static Hamiltonian. 
  
\noindent \emph{Floquet engineering protocol}.-- We now provide the basic steps to Floquet engineer a two-band system, where the desired stroboscopic Hamiltonian is  $H^{{\rm eff}}_{\bs{k}} = h^{\rm eff}_{\bs{k}0} \mathbbm{1} + \bs{h}^{{\rm eff}}_{\bs{k}} \bs{\cdot} \bs{S}$. The protocol is divided into three essential steps: 


\noindent {\bf 1.} \emph{Wei-Norman Ansatz}: Use $U_{\bs{k}}(t) = P_{\bs{k}}(t) e^{-i H^{{\rm eff}}_{\bs{k}} t}$ via Floquet theorem, and apply the Wei-Norman ansatz to construct the micro-motion operator
\be
P_{\bs{k}}(t) = e^{-im_{\bs{k}0}} e^{-im_{\bs{k}+} S_+} e^{-im_{\bs{k}-} S_-} e^{-im_{\bs{k}z} S_z}.
\label{eq:PI(t)-1}
\ee
The function $m_{k0}$ is real, but the other functions $\left(m_{\bs{k}\pm}, m_{\bs{k}z}\right)$ are complex. The explicit time-dependence of the functions $m$ has been suppressed for notational simplicity. Also note that, the last three terms of the above expression are not individually unitary, but their product is unitary which imposes
\begin{eqnarray}
{\rm Im}[m_{\bs{k}z}] &=& \ln\left(1+|m_{\bs{k}+}|^2\right), \nonumber \\ 
m_{\bs{k}-} &=& \frac{m_{\bs{k}+}^*}{1+|m_{\bs{k}+}|^2}.
\end{eqnarray}
The above condition reduces the seven independent parameters (real $m_{\bs{k}0}$ and real and imaginary parts of $m_{\bs{k}\pm ,z}$) to four. We choose $\left(m_{\bs{k}0}, m_{\bs{k}+}, m_{\bs{k}+}^* , m_{\bs{k}z}^{\rm R} \right)$, where $m_{\bs{k}z}^{\rm R} = {\rm Re}[m_{\bs{k}z}]$, as the independent variables and the micro-motion operator reads
\be
P_{\bs{k}}(t) = \frac{e^{-im_{\bs{k}0}}}{\sqrt{1+|m_{\bs{k}+}|^2}} \begin{pmatrix} e^{-\frac{i}{2}m_{\bs{k}z}^{\rm R}} & -i m_{\bs{k}+} e^{\frac{i}{2} m_{\bs{k}z}^{\rm R}}\\ -i m_{\bs{k}+}^* e^{-\frac{i}{2} m_{\bs{k}z}^{\rm R}} & e^{\frac{i}{2} m_{\bs{k}z}^{\rm R}} \end{pmatrix}.\nonumber
\ee

\noindent {\bf 2.} \emph{Transformation matrices}: Consider $\bs{h}^{{\rm eff}}_{\bs{k}}  = \{h^{{\rm eff}}_{\bs{k}-}, h^{{\rm eff}}_{\bs{k}+}, h^{{\rm eff}}_{\bs{k}z}\}$ in $(\pm Z)$ representation. Substituting $U_{k}(t)$ in the TDSE and using Eq. \eqref{eq:equate_coeff}, we obtain $\bs{\mathcal{M}}_1$ and $\bs{\mathcal{M}}_2$ for a given $k$ as
\begin{eqnarray}
\label{eq:transmats}
\bs{\mathcal{M}}_{\bs{k}1} &=& \frac{1}{1+|m_{\bs{k}+}|^2} \begin{pmatrix} 1 & 0 & im_{\bs{k}+}\\ 0 & 1 & -im_{\bs{k}+}^*\\ im_{\bs{k}+}^* & -im_{\bs{k}+} & 1-|m_{\bs{k}+}|^2\end{pmatrix}  \\
\bs{\mathcal{M}}_{\bs{k}2} &=& \frac{1}{1+|m_{\bs{k}+}|^2} \begin{pmatrix} 
e^{-im_{\bs{k}z}^{\rm R}} & -iq_{\bs{k}} m_{\bs{k}+} & im_{\bs{k}+} \\ 
iq_{\bs{k}}^* m_{\bs{k}+}^{*}  & e^{im_{\bs{k}z}^{\rm R}} & -im_{\bs{k}+}^*\\ 
-2 q_{\bs{k}}^*  & -2 q_{\bs{k}}  & 1-|m_{\bs{k}+}|^2 \end{pmatrix}, \nonumber
\end{eqnarray}
with $q_{\bs{k}}=i m_{\bs{k}+} e^{im_{\bs{k}z}^{\rm R}}$. The above form ensures that these matrices are identity at $t=nT$ for $n = 0, 1, 2, \cdots$.

\noindent {\bf 3.} \emph{Driving protocol}: For the Floquet engineering protocol, a bare minimal Hamiltonian is considered as the initial static Hamiltonian $H_0$. For example, here we set $\bs{h} = 0$, i.e. $H_0 = h_0 \mathbbm{1}$. Therefore, using the previous two steps, the driving functions are
\be
\begin{split}
&h_0 + f_{\bs{k}0}(t) = \dot{m}_{\bs{k}0} + \bs{h}^{{\rm eff}}_{\bs{k}0},\\
&\bs{f}_{\bs{k}}(t) =  \bs{\mathcal{M}}_{\bs{k}1} \bs{\cdot} \dot{\bs{m}}_k + \bs{\mathcal{M}}_{\bs{k}2} \bs{\cdot} \bs{h}^{{\rm eff}}_{\bs{k}}.
\end{split}
\label{eq:driving01}
\ee
The transformation matrices are well-defined at all times $t$ (implicit dependence in $m_{\bs{k}}$) and not just stroboscopically [see from Eq.~\eqref{eq:transmats}]. Moreover, the globally well-defined $\bs{\mathcal{M}}_{\bs{k}}\,\forall \bs{k}$ ensures the driving protocol is well-defined for all times. It is worth emphasizing our Floquet engineering protocol is \emph{exact} in the driving frequency $\omega$ and does not require frequency-based perturbative expansions that lead to non-convergent series \cite{Goldman-Dalibard,BW-Oka,Eckardt,EckardtRMP}.

\noindent \emph{Guiding principle to fix the gauge of the micro-motion operator.}-- The gauge for the micro-motion is fixed by choosing $\{m_{\bs{k}0},\bs{{m}_k}\}$ satisfying the boundary conditions at $t=nT$: $e^{-im_{\bs{k}0}\mathbbm{1}}=e^{-im_{\bs{k}+} S_\pm} = e^{-im_{\bs{k}z} S_z}= \mathbbm{1}, \forall n$. This can be achieved in various ways and here we illustrate a physically motivated gauge choice. We first consider the natural choice of a separable form in which each $m_{\bs{k}}(t)$ is a product of momentum and time-dependent functions, such that $\{m_{\bs{k}0}(t),\bs{{m}_k}(t)\} = \{\phi_{\bs{k}0} \mu_0(t), \phi_{\bs{k}+} \mu_+(t), \phi_{\bs{k}+}^* \mu_+^*(t), \phi_{\bs{k}z}^{\rm R} \mu_z^{\rm R}(t)\}$. Furthermore, we set $\phi_{\bs{k}z}^{\rm R} = 1$ and $\phi_{\bs{k}+} = e^{i\bs{k}}$ suggesting that intra sub-lattice hopping is suppressed during the micro-motion and only inter sub-lattice hopping is allowed. Consequently, Eq. \eqref{eq:driving01} simplifies as,
\begin{eqnarray}
f_{\bs{k}0}(t) &=& \phi_{\bs{k}0} \dot{\mu}_0(t) + \bs{h}^{{\rm eff}}_{\bs{k}0} \nonumber \\
\bs{f}_{\bs{k}}(t) &=&  \bs{\tilde{\mathcal{M}}}_{\bs{k}1}\bs{\cdot} \dot{\bs{\mu}}_{\bs{k}}(t) + \bs{\tilde{\mathcal{M}}}_{\bs{k}2}\bs{\cdot}  \bs{h}^{{\rm eff}}_{\bs{k}},
\label{eq:driving02}
\end{eqnarray}
where
\begin{eqnarray}
\bs{\tilde{\mathcal{M}}}_{\bs{k}1} &=& \frac{1}{1+|\mu_{+}(t)|^2} \begin{pmatrix} e^{i\bs{k}} & 0 & i \mu_{+}(t) e^{i\bs{k}} \\ 0 & e^{-i\bs{k}} & i \mu_+^*(t) e^{-i\bs{k}} \\ i\mu_+^*(t) & -i \mu_+(t) & 1-|\mu_+(t)|^2 \end{pmatrix},\nonumber
\label{eq:M-matrix}
\end{eqnarray}
and $\bs{\mathcal{M}}_{\bs{k}2}\rightarrow \bs{\tilde{\mathcal{M}}}_{\bs{k}2}$, defined in Eq.~\eqref{eq:transmats}, with $m_{\bs{k}z}^{\rm R} \rightarrow \mu_z^{\rm R}(t)$ and $m_{\bs{k}+}\rightarrow\mu_+(t)e^{i\bs{k}}$. We now set $\mu_0(t) = a_0 \sin (\omega t)$, $\mu_+(t) = a_+ e^{i\theta} \sin (\omega t)$, and $\mu_z^{\rm R}(t) = p \omega t$ where $p$ is any integer. This choice respects the boundary conditions and ensures the frequency of all the time-dependent functions equals $\omega = 2\pi/ T$. The real amplitudes $\{a_0, a_+\}$, the phase factor $e^{i\theta}$, and the integer $p$ are arbitrary that depend on the physical system as shown below with a specific example. For any arbitrary values of the parameters $a_0,\,a_+,\,\theta$, and $p$, the matrices $\bs{\tilde{\mathcal{M}}}_{\bs{k}1}$ and $\bs{\tilde{\mathcal{M}}}_{\bs{k}2}$ are globally well-defined which ensures the validity of the Wei-Norman ansatz for all $\bs{k}$ and $t$.

\begin{figure}[t]
\includegraphics[width=\columnwidth]{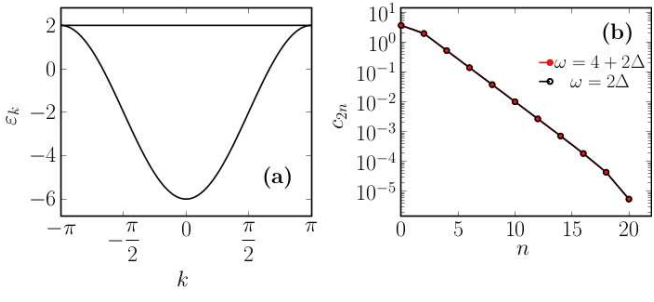}
\caption{(a) Band diagram of the cross-stitch lattice for $\alpha=1.0$ and $\Delta=2.0$. (b) Fourier coefficients of the envelope function for $a_+^2=2.0$.}
\label{fig:band}
\end{figure}

\noindent\emph{Application}.-- We now apply our Floquet engineering protocol to realize the {\it cross-stitch lattice} Hamiltonian $H_{k}^{\rm eff}$, which is a two-band system whose one band is dispersionless (flat) and the other is dispersive~\cite{Flatband01,Flatband02}. In the momentum-space, our target Hamiltonian is $H_{k}^{\rm eff} = h_{k0}^{\rm eff} \mathbbm{1} + h_k^{\rm eff} (S_+ + S_-)$, where $h_{k0}^{\rm eff} = -2\alpha \cos (k)$ and $h_k^{\rm eff} = -(2\alpha\cos (k) + \Delta)$. The energy of the flat band is $\Delta$ and the dispersive band is $-4\alpha\cos (k)  - \Delta$ [see Fig.~\ref{fig:band}(a)].

We choose our initial bare static Hamiltonian $H_{k0} = h_{k0} \mathbbm{1}$, where $h_{k0} = - 2 \alpha \cos (k)$, which describes two uncoupled sub-lattices with each sub-lattice being a $1$D chain with zero onsite energy. The parameter $\alpha$ determines the nearest neighbor hopping strength in each of the sub-lattices. The choice of the simple static Hamiltonian reduces the complexity of the expressions and we obtain $h^{\rm eff}_{k0} = 0$ implying $f_{k0}(t) = \phi_{k0} \dot{\mu}_0(t)$ using Eq.~\eqref{eq:driving02}. As mentioned earlier, the gauge can be fully set with a physical model and hence in this case we have a freedom to set $\phi_0(k) = 0$ and $\theta=0$. Thus, we have $f_{k0}(t) = 0~\forall t$ and the function $\mu_+(t)$ becomes real. 

Using Eq.~\eqref{eq:driving02} and the relations $f_{kx}(t) = 2\, {\rm Re}[f_{k-}(t)]$ and $f_{ky}(t) = - 2\, {\rm Im}[f_{k-}(t)]$, we obtain the driving functions in the $XYZ$ representation as
\begin{eqnarray}
f_{kx}(t) &=& f_{\rm e}(t) \left[a_+\omega \,{\rm C}_{\omega t} {\rm C}_k  + h^{\rm eff}_{k}\,{\rm C}_{p\omega t}  - a_+p \omega\, {\rm S}_{\omega t} {\rm S}_{k} \right.\nonumber \\
&&\left.+ a_+^2 h^{\rm eff}_{k} \,{\rm C}_{2k+p\omega t} {\rm S}^2_{\omega t} \right],\label{eq:drive-kx}\nonumber \\
f_{ky}(t) &=& - f_{\rm e}(t) \left[a_+ \omega\, {\rm C}_{\omega t}{\rm S}_k - h^{\rm eff}_{k}\,{\rm S}_{p\omega t} + a_+p \omega\, {\rm S}_{\omega t} {\rm C}_k \right. \nonumber \\
&&\left. + a_+^2 h^{\rm eff}_{k}\, {\rm S}_{2k+p\omega t}  {\rm S}^2_{\omega t}\right],\label{eq:drive-ky}\nonumber \\
f_{kz}(t) &=& f_{\rm e}(t) \bigg[p \omega \left(1-\frac{a_+^2}{2}\right) + \frac{p \omega a_+^2}{2}  {\rm C}_{2 \omega t} \nonumber \\
&&+ 4 a_+ h^{\rm eff}_{k}\, {\rm S}_{k+p\omega t}{\rm S}_{\omega t}   \bigg], \label{eq:drive-kz}
\end{eqnarray}
where $f_{\rm e}(t) = \left(1+ a_+^2 \, {\rm S}^2_{\omega t}\right)^{-1}$, ${\rm C}_{w}=\cos (w)$ and ${\rm S}_{w}=\sin (w)$. In the above expression, we have two free parameters: a real parameter $a_+$ and an integer $p$. We set these two parameters such that each of the driving function does not have any static part. First, we set $a_+^2 = 2$ which removes the first term of the driving function $f_{kz}(t)$. Next we set $p=3$, which is the minimal integer that ensures absence of any static term in the driving protocols~\cite{note}. The above protocol in lattice space turns out to be \emph{local} involving only the next-to-next nearest neighbors, ensuring experimental feasibility (see supplementary~\cite{SI}).  
\begin{figure}[b]
\includegraphics[width=\columnwidth]{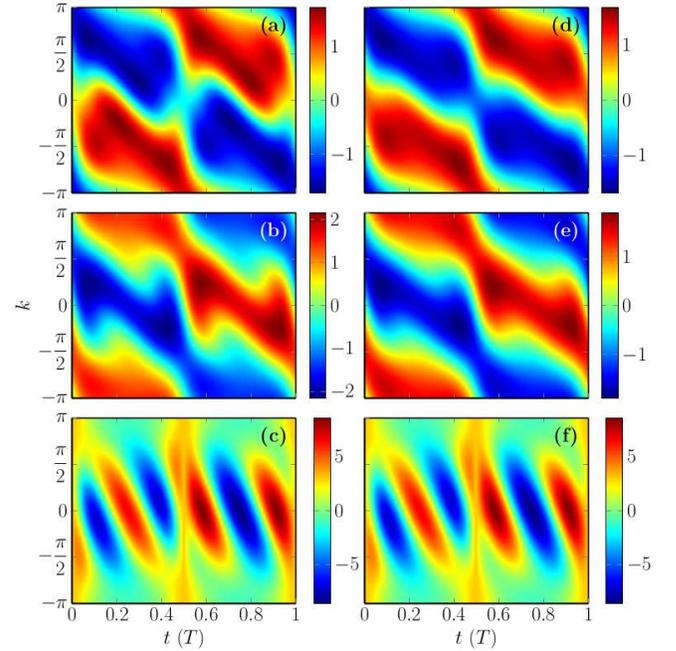}
\caption{Density plot of the driving functions $f_{kx}(t)$ (a,d), $f_{ky}(t)$ (b,e), $f_{kz}(t)$ (c,f) are plotted as a function of momentum $k$ and time $t$ for $\omega = 4 + 2 \Delta = 8$ (a-c) and $\omega = 2 \Delta = 4$ (d-f).} 
\label{fig:drive}
\end{figure}

We consider two moderate (same order of the band gap) cases of the driving frequency: $\omega = 4 + 2 \Delta = 8$ and $\omega = 2 \Delta = 4$. For these two cases, the Fourier coefficients are shown in Fig.~\ref{fig:band}(b) as a function of the coefficient indices. For both frequencies, the odd coefficients $c_{2n+1}$ are zero and the even coefficients $c_{2n}$ fall exponentially with $n$. Therefore, the envelope can be realized with high accuracy considering only a few even harmonics. Figure~\ref{fig:drive} illustrates all the driving functions, given by Eq.~\eqref{eq:drive-kz}, for driving frequency $\omega = 4 + 2 \Delta = 8$ [panels (a)-(c)] and $\omega = 2 \Delta = 4$ [panels (d)-(f)]. Clearly, these are not simple functions having sine or cosine periodicity in time $t$ as typically considered in the literature.

\noindent \emph{Conclusion}.-- We have introduced a Floquet engineering protocol applicable to systems whose Hamiltonians have an underlying Lie-algebraic structure. A large number of physically relevant interacting and noninteracting models in any dimension fall into this class of systems. In our formalism, we have applied the Wei-Norman ansatz \cite{WeiNorman-JMP} to the micro-motion part of the Floquet dynamics, and from that, we have prescribed how to design a driving protocol to reach the desired system starting from a given simple static Hamiltonian. We have explicitly solved the case of two-band systems that obey the unsolvable $SU(2)$ algebra and described a guiding principle to fix the gauge of the micro-motion operator. We then illustrate our idea by stroboscopically realizing the cross-stitched model's flat and dispersive band diagram~\cite{Flatband01,Flatband02}.

Our formulation does not rely on any perturbative expansions and is exact. The main idea is to provide a recipe to design a driving protocol on a simple static Hamiltonian such that a desired stroboscopic Hamiltonian is obtained for \emph{any} driving frequency. Unlike previous works that rely on a specific form of the driving (sine or cosine), the exactness of our approach allows us to engineer the functional form of the drive. Even though we tackled generic condensed matter setup of systems having two-energy bands, our formalism can be easily adapted for any two-level system with a driving protocol that is a generalization to those studied in Refs. \cite{Barnes-DasSarma,Barnes,Nori}.


In principle, the formalism presented here can be applied to multi-band systems (see supplementary~\cite{SI} for a discussion on three-bands systems) to Floquet engineer technologically relevant materials like higher-order topological insulator (HOTI) \cite{HOTI,FHOTI-1,FHOTI} or reproduce $\mathbbm{Z}_2$ lattice gauge theory in cold atom setup \cite{LGT}.

\begin{acknowledgments}
\noindent \emph{Acknowledgments}.--
This research was supported by the Institute for Basic Science in Korea (IBS-R024-Y2). JNB acknowledges financial support from DST-SERB, India through the Core Research Grant CRG/2020/001701.
\end{acknowledgments}


\begin{widetext}
\vspace{0.5cm}

\begin{center}
{\large\bf Supplementary Information}
\end{center}

\vspace{-0.5cm}
\begin{appendix}

\section{General formalism}
In this section we present the details of the general formalism presented in the main text. We begin with a general form of a time-dependent Hamiltonian having a period $T$ that reads,
\be
H(t) = H_0 + V(t),~V(t+T) = V(t),
\label{eq:Ham01}
\ee
where $H_0$ is the time-independent static part and $V(t)$ is the periodic driving. Utilizing operators $A_{\alpha}$ that form a closed Lie algebra of dimension $N$, we can express the Hamiltonian as
\be
H_0 = h_0 \mathbbm{1} + \sum_{\alpha = 1}^N h_\alpha\, A_\alpha = h_0 \mathbbm{1} + \bs{h \cdot A} ~~{\rm and}~~V(t) = f_0(t) \mathbbm{1} + \sum_{\alpha = 1}^N f_\alpha(t) A_\alpha = f_0(t)\, \mathbbm{1} + \bs{f}(t) \bs{\cdot A},
\label{eq:Ham02}
\ee
where $h$'s are functions of the system parameters, the driving functions $f(t)$'s are functions of both time $t$ and system parameters. The algebraic structure of $A$'s is governed by the structure constants $\Lambda_{\alpha\beta}^\gamma$ which are defined in the main text via Eq.~(6). The time-periodic condition of $V(t)$ implies that all the driving functions are also time-periodic. If $U(t)$ is the corresponding time-evolution operator, then this operator will satisfy the time-dependent Schr\"odinger equation (TDSE) with $\hbar=1$,
\be
i \frac{dU(t)}{dt} = H(t)\, U(t),
\label{eq:TDSE}
\ee
whose solution is
\be
U(t) = \mathcal{T} \exp\left\{- i\, \int_0^t H(t^\prime)\,dt^\prime\right\},
\ee 
where $\mathcal{T}$ is the time-ordering operator. The Floquet theorem suggests that the above time-evolution operator can always be written as a product two unitary operators: operator $P(t)$ that describes the short-time dynamics within one period with $P(t+T) = P(t)$; the other operator describes the long time dynamics governed by an effective static Hamiltonian $H_F$ called the `Floquet Hamiltonian'. Therefore, we have
\be
U(t) = P(t)\, e^{- i H_F t}.
\ee
Following the initial condition $U(0) = \mathbbm{1}$, we get $P(0) = \mathbbm{1}$. Moreover, the time-periodic property of $P(t)$ suggests that $P(nT) = \mathbbm{1}$ for any arbitrary positive integer $n$. If one observes the dynamics of a particle stroboscopically at every time interval $nT$, then that dynamics is effectively governed by the static Hamiltonian $H_F$. From the Floquet engineering perspective, this effective static Hamiltonian is also the desired one which {\it Floquet engineers} want to obtain by designing a driving protocol. Therefore, in the remaining part of this supplementary information, we denote the Floquet Hamiltonian $H_F$ by the effective Hamiltonian $H^{\rm eff}$. Since this Hamiltonian is obtained from a time-dependent Hamiltonian with underlying Lie algebraic structure, the general form of $H^{\rm eff}$ will also be a linear combination of $\{\mathbbm{1}, \, A_\alpha\}$ of the form
\be
H^{\rm eff} = h^{\rm eff}_0 \mathbbm{1} + \sum_{\alpha=1}^N h^{\rm eff}_{\alpha} \, A_\alpha = h^{\rm eff}_0 \mathbbm{1} + \bs{h}^{\rm eff} \bs{\cdot A}.
\label{eq:Heff}
\ee

Substituting the Floquet form of the time-evolution operator in the TDSE, given by Eq.  \eqref{eq:TDSE}, we obtain
\be
\left[i \frac{dP(t)}{dt} + P(t) H^{\rm eff}\right] e^{-i H^{\rm eff}t}  = H(t)\, U(t).
\label{eq:TDSE-01}
\ee
The idea of Floquet engineering implies that the form of the evolution operator corresponding to the long time dynamics is known, but the micro-motion part is unknown. Since the Hamiltonian has an underlying Lie algebraic structure, we can apply the Wei-Norman ansatz~\cite{WeiNorman-JMP,WeiNorman-PAMS} to the micro-motion operator. According to this ansatz, we can write
\be
P(t) = e^{-i m_0(t)} \prod_{\alpha = 1}^N e^{-i m_\alpha(t) A_\alpha}.
\ee
From the condition $P(nT) = \mathbbm{1}$, we get the following conditions
\be
e^{-i m_0(nT)} = 1~~{\rm and}~~e^{-i m_\alpha(nT) A_\alpha} = \mathbbm{1}\,\, \forall\,\, \alpha ~{\rm and}~n.
\ee
Substituting the above Wei-Norman form of $P(t)$ in Eq. \eqref{eq:TDSE-01}, then the first term at the left hand side will become
\be
\begin{split}
\label{eq:generalTDSE1}
&i \frac{dP(t)}{dt} e^{-i H^{\rm eff}t} = \left[\dot{m}_0(t)\, P(t) + \dot{m}_1(t)\, e^{-i m_0(t)} \underbrace{A_1 \left\{\prod_{\alpha = 1}^N e^{-i m_\alpha(t) A_\alpha}\right\}}_{= A_1 P(t)}\right.\\
&+ \dot{m}_2(t)\, e^{-i m_0(t)}\, \underbrace{e^{-i m_1(t) A_1} A_2}_{\rm applying\,\,BCH} \left\{\prod_{\alpha = 2}^N e^{-i m_\alpha(t) A_\alpha} \right\} + \cdots \\
&+ \dot{m}_\beta(t)\, e^{-i m_0(t)} \underbrace{\left\{\prod_{\alpha = 1}^{\beta - 1} e^{-i m_\alpha(t) A_\alpha}\right\} A_\beta}_{\rm applying\,\,BCH} \left\{\prod_{\alpha^\prime = \beta}^N e^{-i m_{\alpha^\prime}(t) A_{\alpha^\prime}}\right\} + \cdots \\
& \left.  + \, \dot{m}_N(t)\, e^{-i m_0(t)} \underbrace{\left\{\prod_{\alpha = 1}^{N - 1} e^{-i m_\alpha(t) A_\alpha}\right\} A_N}_{\rm applying\,\,BCH}\, e^{-i m_N(t) A_N}\right] e^{-i H^{\rm eff}t}.
\end{split} 
\ee
In the above expression, the first and the second terms at the right side are equal to $\dot{m}_0(t)\, U(t)$ and $\dot{m}_1(t)\, A_1\, U(t)$, respectively. If we apply Baker-Campbell-Hausdorff (BCH) formula \cite{WeiNorman-JMP} to the expression $e^{-i m_1(t) A_1} A_2$ of the third term, then we can {\it push} the exponential operator to the right and the third term will take the form 
\[{\rm third\,\,term} = \dot{m}_2(t) \left\{\sum_{\gamma = 1}^N \mathcal{M}_{1\gamma}^{(3)}(m_1) A_\gamma\right\} U(t), \]
where the coefficients $\mathcal{M}_{1 \gamma}^{(3)}(m_1)$ are in general nonlinear functions of $m_1(t)$. The term is a linear function of the operators $A_\gamma$'s because these operators form a closed Lie algebra. Similarly, for the general $(\beta+1)$th term of the above expression, we can push all the exponential operators in the following expression  
\[\left\{\prod_{\alpha = 1}^{\beta - 1} e^{-i m_\alpha(t) A_\alpha}\right\} A_\beta\]
to the right by applying the BCH formula $(\beta-1)$ times. Then the general $(\beta+1)$th term will be of the following form
\[(\beta+1){\rm~term} = \dot{m}_\beta(t) \left\{\sum_{\gamma=1}^N \mathcal{M}_{1\gamma}^{(\beta+1)}(m_1, m_2, \cdots, m_{\beta-1})\, A_\gamma\right\} U(t).\]
Here again the coefficients $\mathcal{M}_{1 \gamma}^{(\beta+1)}(m_1, m_2, \cdots, m_{\beta-1})$ are nonlinear functions of $(m_1(t), m_2(t), \cdots, m_{\beta-1}(t))$ and the linearity of the operators $A_\gamma$'s is due to the closed Lie algebraic structure as mentioned above. 
Therefore, using all above relations in Eq.~\eqref{eq:generalTDSE1} we obtain,
\be
\label{eq:M1}
i \frac{dP(t)}{dt} e^{-i H^{\rm eff}t} = \left[\dot{m}_0(t)\, \mathbbm{1} + \underbrace{\dot{\bs{m}}(t) \bs{\cdot} \bs{\mathcal{M}}_1(t)^{\rm T}}_{=\, \bs{\zeta}(\bs{m},\, \bs{\dot{m}})} \bs{\cdot} \bs{A}\right] U(t),
\ee
where $\bs{m} = \{m_{\alpha}(t)\}$ is a $N$-dimensional vector whose components are time-dependent functions $m_\alpha(t)$, $\bs{A} =\{A_{\alpha}\} $ is a vector whose each element is the operator $A_\alpha$, the elements of the matrix $\bs{\mathcal{M}_1}(t)$ are in general nonlinear functions of $\bs{m}(t)$.

We now consider the second term on the left-hand side of Eq.~\eqref{eq:TDSE-01},
\be
\label{eq:M2}
P(t)\, H^{\rm eff}\, e^{-i H^{\rm eff}t} = \left[P(t)\, H^{\rm eff}\, P^\dagger(t)\right]\, U(t) = \left[h^{\rm eff}_0 \mathbbm{1} + \sum_{\alpha = 1}^N h^{\rm eff}_{\alpha} \Bigl\{P(t)\, A_\alpha\, P^\dagger(t)\Bigr\}\right]\,U(t).
\ee
Again due to the closed Lie algebra of $\{A_\alpha\}$, applying the BCH formula, we have
\be
\sum_{\alpha = 1}^N h^{\rm eff}_{\alpha} \Bigl\{P(t)\, A_\alpha\, P^\dagger(t)\Bigr\} = \underbrace{\bs{h}^{\rm eff} \bs{\cdot} \bs{\mathcal{M}}_2 (t)^{\rm T}}_{=\,\bs{\xi}(\bs{h}^{\rm eff}, \bs{m})} \bs{\cdot} \bs{A}.
\ee
Therefore, Eq.~\eqref{eq:TDSE-01} becomes
\be
\left[\dot{m}_0(t)\, \mathbbm{1} + \dot{\bs{m}} \bs{\cdot} \bs{\mathcal{M}}_1(t)^{\rm T} \bs{\cdot} \bs{A} + h^{\rm eff}_0 \mathbbm{1} + \bs{h}^{\rm eff} \bs{\cdot} \bs{\mathcal{M}}_2^{\rm T} (t) \bs{\cdot} \bs{A} \right] U(t) = H(t) U(t).
\ee
Using the Hamiltonian $H(t)$ given by Eqs.~\eqref{eq:Ham01}-\eqref{eq:Ham02} and equating the coefficients of $\mathbbm{1}$ and $A_\alpha$'s from both sides, we obtain
\be
h_0 + f_0(t) = \dot{m}_0(t) + h^{\rm eff}_0 ~~{\rm and}~~\bs{h} + \bs{f}(t) = \bs{\mathcal{M}}_1 (t) \bs{\cdot} \dot{\bs{m}} + \bs{\mathcal{M}}_2 (t) \bs{\cdot} \bs{h}^{\rm eff}.
\ee 
which matches Eq. (16) from the main text.

\section{Formalism for the two-bands case: Floquet Engineering Protocol}

The two-band Hamiltonians are represented by the operators that follow $SU(2)$ algebra. Here we represent the Hamiltonian in terms of the operators $(\mathbbm{1}, S_\pm, S_z)$, where $S_\pm = S_x \pm i S_y$ and $\mathbbm{1}$ is a $2\times 2$ identity matrix. Here, $2\bs{S} = \bs{\sigma}$ where $\bs{\sigma} = (\sigma_x, \sigma_y, \sigma_z)$ are spin-$1/2$ Pauli matrices. Following the Wei-Norman ansatz, the corresponding micro-motion operator for each $\bs{k}$ can be written as
\be
P_{\bs{k}}(t) = e^{-im_{{\bs k}0}(t)\mathbbm{1}} e^{-i m_{{\bs k}+}(t) S_+} e^{-i m_{{\bs k}-}(t) S_-} e^{-i m_{{\bs k}z}(t) S_z}.
\label{eq:Wei-Norman}
\ee
Since, $S_\pm$ are not Hermitian, then $m_{\bs{k}\pm}(t)$ are complex and $e^{-i m_{\bs{k}\pm}(t) S_\pm}$ is not unitary, but over all the operator $P(t)$ is unitary. Moreover, even though $S_z$ is a Hermitian operator, we still have to consider $m_{\bs{k}z}(t)$ to be complex. This is because, for the real $m_{\bs{k}z}(t)$, we cannot find any pair of complex numbers $m_{\bs{k}\pm}(t)$ except zeros, such that $P(t)$ will be unitary. From the unitary property of $P(t)$, we have shown in the main text that $m_{{\bs k}-}$ and ${\rm Im}[m_{{\bs k}z}]$ are not independent functions and both of these can be expressed in terms of $m_{{\bs k}+}$ (see Eq. (14) of the main text). 

We now have to construct $\bs{\mathcal{M}}_1(t)$ and $\bs{\mathcal{M}}_2(t)$ matrices for each momentum value $\bs{k}$ to determine the driving protocol. We denote these matrices as $\bs{\mathcal{M}_{k}}_1(t)$ and $\bs{\mathcal{M}_{k}}_2(t)$, respectively. The matrix $\bs{\mathcal{M}_{k}}_1(t)$ will be constructed from the time derivative of $P_{\bs{k}}(t)$, Eq.~\eqref{eq:M1}, using the Wei-Norman form given in Eq. \eqref{eq:Wei-Norman}. Therefore, we obtain 
\be
\begin{split}
i \frac{dP_{\bs{k}}(t)}{dt} &= \left[\dot{m}_{\bs{k}0} \mathbbm{1} + \underbrace{\left\{\dot{m}_{\bs{k}+} + m_{\bs{k}+}^2 \dot{m}_{\bs{k}-} + i m_{\bs{k}+} (1 - m_{\bs{k}+} m_{\bs{k}-}) \dot{m}_{\bs{k}z} \right\}}_{=\, C_+} S_+ \right.\\ &+ \left. \underbrace{\left(\dot{m}_{\bs{k}-} - im_{\bs{k}-} \right)}_{=\, C_-} S_- + \underbrace{\left\{\dot{m}_{\bs{k}z} (1-2m_{\bs{k}+} m_{\bs{k}-}) - 2 i m_{\bs{k}+}\dot{m}_{\bs{k}-} 
\right\}}_{=\, C_z} S_z\right] P_{\bs{k}}(t).
\end{split}
\label{eq:deriv_Pk}
\ee  
Above, just like the main text we suppress the explicit time dependence in the functions $m$ for notational simplicity. We now express above equation in terms of two independent variables $m_{\bs{k}+}$ and the real part of $m_{\bs{k}z}$, i.e., $m_{\bs{k}z}^{\rm R}$. Therefore, we replace $m_{\bs{k}-}$ and the imaginary part of $m_{\bs{k}z}$, i.e., $m_{\bs{k}z}^{\rm I}$, by the following expression given in the main text (see Eq.~(14) in the main text),
\be
m_{\bs{k}-} = \frac{m_{\bs{k}+}^*}{1+|m_{\bs{k}+}|^2}~~{\rm and}~~m_{\bs{k}z}^{\rm I} = \ln(1+|m_{\bs{k}+}|^2),
\label{eq:dep_var}
\ee
where $m_{\bs{k}+}^*$ is the complex conjugate of $m_{\bs{k}+}$. Subsequently, the time derivative of these functions are
\be
\dot{m}_{\bs{k}-} = \frac{\dot{m}_{\bs{k}+}^* - m_{\bs{k}+}^{*2} \dot{m}_{\bs{k}+}}{\left(1+|m_{\bs{k}+}|^2\right)^2}~~{\rm and}~~\dot{m}_{\bs{k}z}^{\rm I} = \frac{m_{\bs{k}+} \dot{m}_{\bs{k}+}^* + m_{\bs{k}+}^* \dot{m}_{\bs{k}+}}{1+|m_{\bs{k}+}|^2}. 
\label{eq:deriv_dep_var}
\ee
Using the relations obtained in Eqs. \eqref{eq:dep_var} and \eqref{eq:deriv_dep_var}, we obtain the coefficients of Eq. \eqref{eq:deriv_Pk} as
\be
\begin{split}
C_+ &= \frac{1}{1+|m_{\bs{k}+}|^2} (\dot{m}_{\bs{k}+} + i m_{\bs{k}+} \dot{m}_{\bs{k}z}^{\rm R}),~~C_- = C_+^*,~~{\rm and} \\C_z &= \frac{1}{1+|m_{\bs{k}+}|^2} \left[i m_{\bs{k}+}^* \dot{m}_{\bs{k}+} - i m_{\bs{k}+} \dot{m}_{\bs{k}+}^* + (1- |m_{\bs{k}+}|^2)\, \dot{m}_{\bs{k}z}^{\rm R} \right].
\end{split}
\label{eq:coeffs}
\ee
Then from Eq. \eqref{eq:deriv_Pk}, equating the coefficients of $\{S_\pm,\, S_z\}$ from the both sides of the TDSE, we get $\bs{\mathcal{M}_{k}}_1$ given by Eq. (15) of the main text. 

In order to derive the matrix $\bs{\mathcal{M}_{k}}_2$ we have to calculate $P_{\bs{k}}(t) H^{\rm eff}_{\bs{k}} P_{\bs{k}}^\dagger(t)$, Eq.~\eqref{eq:M2}, where $H^{\rm eff}_{\bs{k}} = P_{\bs{k}}(t) H^{\rm eff}_{\bs{k}} P_{\bs{k}}^\dagger(t) + \bs{h}^{\rm eff}_{\bs{k}} \bs{\cdot} \bs{S}$. In the $(S_x, S_y, S_z)$ and $(S_\pm, S_z)$ representation, $H^{\rm eff}_{\bs{k}}$ is
\be
H^{\rm eff}_{\bs{k}} = h^{\rm eff}_{\bs{k}0} \mathbbm{1} + h_{{\bs k}x} S_x + h_{{\bs k}y} S_y + h_{{\bs k}z} S_z = h^{\rm eff}_{\bs{k}0} \mathbbm{1} + h_{{\bs k}-} S_+ + h_{{\bs k}+} S_- + h_{{\bs k}z} S_z,
\label{eq:Heff-2b}
\ee
where $2h_{{\bs k}\pm} = h_{{\bs k}x} \pm i h_{{\bs k}y}$. Here we consider the $\pm Z$ representation to obtain,
\be
P_{\bs{k}}(t) H^{\rm eff}_{\bs{k}} P_{\bs{k}}^\dagger(t) = h^{\rm eff}_{\bs{k}0} \mathbbm{1} + P_{\bs{k}}(t)\, \left(h_{{\bs k}-} S_+ + h_{{\bs k}+} S_- + h_{{\bs k}z} S_z\right)\,  P_{\bs{k}}^\dagger(t).
\label{eq:PHeffPdag}
\ee
We have found that
\be
\begin{split}
P_{\bs{k}}(t)\, S_+\, P_{\bs{k}}^\dagger(t) &= \frac{e^{-im_{\bs{k}z}^{\rm R}}}{1+|m_{\bs{k}+}|^2} \left(S_+ + m_{\bs{k}+}^{*2} S_- + 2 i m_{\bs{k}+}^* S_z\right),\\
P_{\bs{k}}(t)\, S_-\, P_{\bs{k}}^\dagger(t) &= \frac{e^{im_{\bs{k}z}^{\rm R}}}{1+|m_{\bs{k}+}|^2} \left(m_{\bs{k}+}^2 S_+ + S_-  - 2 i m_{\bs{k}+} S_z\right),\\
P_{\bs{k}}(t)\, S_z\, P_{\bs{k}}^\dagger(t) &= \frac{1}{2(1+|m_{\bs{k}+}|^2)} \left[2im_{\bs{k}+} S_+ - 2im_{\bs{k}+}^* S_- + 2(1-|m_{\bs{k}+}|^2) S_z \right].
\end{split}
\label{eq:PSPdag}
\ee
The above relations can be derived in two ways: (1) Applying the BCH formula multiple times. This is a cumbersome approach. However, it is a very general method that can be applied for any closed algebra irrespective of its representation in any dimension. (2) A straightforward way is to explicitly write down the matrix representation of $P_{\bs{k}}(t)$ and $S$-matrices, then calculate matrix multiplication of three matrices to obtaining each of the three relations. Substituting the results obtained in Eqs. \eqref{eq:deriv_Pk} and \eqref{eq:PSPdag} in the TDSE $idU_{\bs{k}}(t)/dt = H_{\bs{k}}(t) U_{\bs{k}}(t)$, and equating the coefficients of $(\mathbbm{1}, S_\pm, S_z)$ from the both sides,  we get the relations given by Eqs. (16) and (17) of the main text.

\section{Real space representation of the driving protocol: Eq. (18) of the main text}

In the main text, we have illustrated how to apply our proposed Floquet engineering protocol to realize a desired two bands system. There we presented the driving protocol in the momentum space. We are presenting the same driving protocol in the real or lattice space. Here our goal is to show that we only require {\it spatially local} driving to realize the desired energy bands.
 
For the two-bands case, we can assume the system is consisted of two sub-lattices. This implies each lattice site has the contribution of two orbitals ($A$ and $B$) or a dimer. In the momentum space ($k$-space), we define the annihilation operators $\tilde{c}_{A,k}$ and $\tilde{c}_{B,k}$, where the first operator annihilates a particle in orbital $A$ having momentum $k$, and the later operator annihilates a particle in orbital $B$ having the same momentum. Analogously we can define the creation operators $\tilde{c}_{A,k}^\dagger$ and $\tilde{c}_{B,k}^\dagger$, where these operators have their usual meaning. If we define $\tilde{\Psi}_k = \left(\tilde{c}_{A,k} ~~ \tilde{c}_{B,k}\right)^{\rm T}$, then the driven part of the {\it full} Hamiltonian can be written as
\be
V(t) = \sum_k \tilde{\Psi}_k^\dagger V_k(t) \tilde{\Psi}_k,~~{\rm where}~~V_k(t) = f_{kx}(t) S_x + f_{ky}(t) S_y + f_{kz}(t) S_z, ~{\rm since}~f_{k0}(t) = 0\,\,\forall\,t.
\label{eq:Vt}
\ee  
Here assume the periodic boundary condition, therefore the momentum $k$ can have discrete values, i.e., $k_n = 2\pi n/L$. The driving functions $f_{k \alpha}(t)$ where $\alpha = x,\,y,\,z$ are given in the main text.

We now analyze term by term of Eq. \eqref{eq:Vt} in the lattice space. 
\be
\begin{split}
{\rm 1st~ term} &= \sum_k f_{kx}(t) \tilde{\Psi}_k^\dagger S_x \tilde{\Psi}_k\\ 
&= f_e(t) \left[\left(a_+ \omega C_{\omega t} - 2 \alpha\right) \sum_k \cos (k)\, \tilde{\Psi}_k^\dagger S_x \tilde{\Psi}_k - a_+ p \omega S_{\omega t} \sum_k \sin (k)\, \tilde{\Psi}_k^\dagger S_x \tilde{\Psi}_k \right.\\ 
&- \left. 2 a_+^2 \alpha S_{\omega t}^2 \sum_k \left(C_{p\omega t} \cos (k) \cos (2k) + S_{p\omega t} \cos ( k) \sin (2k)\right) \,\tilde{\Psi}_k^\dagger S_x \tilde{\Psi}_k \right]\\
&= f_e(t) \left[\left(a_+ \omega C_{\omega t} - 2 \alpha\right) \sum_k \cos (k)\, \tilde{\Psi}_k^\dagger S_x \tilde{\Psi}_k - a_+ p \omega S_{\omega t} \sum_k \sin (k)\, \tilde{\Psi}_k^\dagger S_x \tilde{\Psi}_k \right.\\
&\left. - a_+^2 \alpha S_{\omega t}^2 \left\{ C_{p \omega t} \sum_k \left(\cos (3k) + \cos (k)\right) \tilde{\Psi}_k^\dagger S_x \tilde{\Psi}_k + S_{p \omega t} \sum_k \left(\sin (3k) + \sin (k)\right) \tilde{\Psi}_k^\dagger S_x \tilde{\Psi}_k \right\}\right],
\end{split}
\label{eq:1st}
\ee
where $C_w = \cos (w)$ and $S_w = \sin (w)$, and we set in the main text $a_+^2 = 2$ and $p=3$.
Similarly, we find that
\be
\begin{split}
{\rm 2nd~ term} &= \sum_k f_{ky}(t) \tilde{\Psi}_k^\dagger S_y \tilde{\Psi}_k\\
&= - f_e(t) \left[ a_+ \omega C_{\omega t} \sum_k \sin (k)\, \tilde{\Psi}_k^\dagger S_y \tilde{\Psi}_k + \left(2\alpha - a_+ p \omega S_{\omega t}\right) \sum_k \cos (k)\, \tilde{\Psi}_k^\dagger S_y \tilde{\Psi}_k \right.\\
&\left. - a_+^2 \alpha S_{\omega t}^2 \left\{ C_{p \omega t} \sum_k \left(\sin (3k) + \sin (k)\right) \tilde{\Psi}_k^\dagger S_y \tilde{\Psi}_k + S_{p \omega t} \sum_k \left(\cos (3k) + \cos (k)\right) \tilde{\Psi}_k^\dagger S_y \tilde{\Psi}_k\right\} \right]
\end{split} 
\label{eq:2nd}
\ee
and
\be
\begin{split}
{\rm 3rd~ term} &= \sum_k f_{kz}(t) \tilde{\Psi}_k^\dagger S_z \tilde{\Psi}_k\\
&= f_e(t) \left[\left(\frac{1}{2} p \omega a_+^2 - 4a_+ \alpha S_{\omega t} S_{p \omega t}\right) \sum_k \tilde{\Psi}_k^\dagger S_z \tilde{\Psi}_k\right.\\ &\left. - 4 a_+ \alpha S_{\omega t} \left\{ C_{p \omega t} \sum_k \sin (2k)\, \tilde{\Psi}_k^\dagger S_z \tilde{\Psi}_k  + S_{p \omega t} \sum_k \cos (2k)\, \tilde{\Psi}_k^\dagger S_z \tilde{\Psi}_k \right\} \right].
\end{split}
\label{eq:3rd}  
\ee
We notice above that the form of all the terms are either
\be
F(t) \sum_k \cos (mk) \, \tilde{\Psi}_k^\dagger S_\alpha \tilde{\Psi}_k ~{\rm with}~m = 0, 1, 2, 3~~{\rm or}~~ F(t) \sum_k  \sin (mk)\, \tilde{\Psi}_k^\dagger S_\alpha \tilde{\Psi}_k ~{\rm with}~m = 1, 2, 3,
\label{eq:general_form}
\ee 
where $F(t)$ is different time-dependent functions and $\alpha = x, y, z$. We can present $V(t)$ in the lattice space by transforming the creation and the annihilation operators from the momentum space to the lattice space, i.e., $(\tilde{c}_{A,k},\, \tilde{c}_{B,k}) \rightarrow (c_{A,n},\, c_{B,n})$. This transformation is defined by the inverse Fourier transformation as:
\be
\tilde{c}_{A/B, k} = \frac{1}{\sqrt{L}} \sum_n c_{A/B, n} e^{i k n}~~{\rm and}~~\tilde{c}_{A/B, k}^\dagger = \frac{1}{\sqrt{L}} \sum_n c_{A/B, n}^\dagger e^{-i k n},
\ee
where $L$ is the number of sites in each of the sub-lattices. Since, we are interested only in the derivation of the lattice space representation of the terms given in Eq. \eqref{eq:general_form}, we are ignoring the time-dependent functions $F(t)$. For $\alpha = x$, we have
\be
\begin{split}
\sum_k \cos (mk)\, \tilde{\Psi}_k^\dagger S_x \tilde{\Psi}_k &= \frac{1}{2} \sum_k \cos (mk) \left(\tilde{c}_{A,k}^\dagger \tilde{c}_{B,k} + \tilde{c}_{B,k}^\dagger \tilde{c}_{A,k}\right)\\
&= \frac{1}{4} \sum_k \left(e^{imk} + e^{-imk}\right)\,\left[ \frac{1}{L} \sum_{n, n^\prime} c_{A,n}^\dagger c_{B,n^\prime} e^{i k (n-n^\prime)} + \frac{1}{L} \sum_{n, n^\prime} c_{B,n}^\dagger c_{A,n^\prime} e^{i k (n-n^\prime)}\right]\\
&= \frac{1}{4} \sum_{n, n^\prime} \left(c_{A,n}^\dagger c_{B,n^\prime} + c_{B,n}^\dagger c_{A,n^\prime}\right) \underbrace{\frac{1}{L} \sum_k e^{i k (n+m-n^\prime)}}_{=\, \delta_{n^\prime, n+m}} +\, {\rm H.c.}\\
&= \frac{1}{4} \sum_n \left(c_{A,n}^\dagger c_{B,n+m} + c_{B,n}^\dagger c_{A,n+m}\right) + \, {\rm H.c.}
\end{split}
\ee
Similarly, we derive
\be
\begin{split}
\sum_k \sin (mk)\, \tilde{\Psi}_k^\dagger S_x \tilde{\Psi}_k &= -\frac{i}{4} \sum_n \left(c_{A,n}^\dagger c_{B,n+m} + c_{B,n}^\dagger c_{A,n+m}\right) + \, {\rm H.c.}\\
\sum_k \cos (mk)\, \tilde{\Psi}_k^\dagger S_y \tilde{\Psi}_k &= \frac{i}{4} \sum_n \left(c_{A,n}^\dagger c_{B,n+m} - c_{B,n}^\dagger c_{A,n+m}\right) + \, {\rm H.c.}\\
\sum_k \sin (mk)\, \tilde{\Psi}_k^\dagger S_y \tilde{\Psi}_k &= \frac{1}{4} \sum_n \left(c_{A,n}^\dagger c_{B,n+m} - c_{B,n}^\dagger c_{A,n+m}\right) + \, {\rm H.c.}\\
\sum_k \cos (mk)\, \tilde{\Psi}_k^\dagger S_z \tilde{\Psi}_k &= \frac{1}{4} \sum_n \left(c_{A,n}^\dagger c_{A,n+m} - c_{B,n}^\dagger c_{B,n+m}\right) + \, {\rm H.c.}\\
\sum_k \sin (mk)\, \tilde{\Psi}_k^\dagger S_z \tilde{\Psi}_k &= \frac{i}{4} \sum_n \left(c_{A,n}^\dagger c_{A,n+m} - c_{B,n}^\dagger c_{B,n+m}\right) + \, {\rm H.c.}
\end{split}
\ee
In the above expressions, the maximum value of $m=3$. Hence, the maximum hopping range is from $n$-th site to $(n+3)$-th site, i.e. from one site to its next-to-next nearest neighbor (NNNN) hopping. This suggests that our driving protocol demands only short range hopping, which is experimentally easily realizable.




\section{Case of the Lie algebraic quantum interacting systems in 1D and 2D}

The Floquet engineering protocol proposed in the Letter is applicable to all quantum systems of any dimension provided the Hamiltonian of systems have any underlying closed Lie algebra. In the main text, we have applied our protocol to realize a 1D non-interacting system. We can straightforwardly extend this formalism to any Lie algebraic higher dimensional systems. For example, graphene is a 2D material. In the tight-binding limit, the Hamiltonian of graphene can be expanded as a linear combination of the Pauli pseudo-spin matrices which form $SU(2)$ algebra. Therefore, graphene comes under the purview of our formalism whose bands can be engineered.

The mean-field theory is a standard formulation of studying many interacting systems. In the mean-field limit, interacting particles in a system can be replaced by non-interacting quasi-particles which are coupled to a {\it well-defined} effective background field \cite{bernevig-book}. The theory of conventional ($s$-wave) and the so called unconventional ($p$-wave) superconductors are prominent examples of interesting physical systems which are studied under the mean-field approximation \cite{bernevig-book}. Moreover, these systems are also represented as a linear combination of the Pauli matrices (or the product of Pauli matrices) in the momentum space. Therefore, they also have underlying closed Lie algebra. Here, we discuss these systems following Ref. \cite{bernevig-book}.     

\subsubsection{Conventional ($s$-wave) superconductors}

The mean-field quasi-particle formulation of the conventional $s$-wave superconductors is described by the Bogoliubov-de-Gennes (BdG) theory \cite{de-Gennes,bernevig-book}. The Hamiltonian of this class of systems has two parts: a pure metallic part and a (Cooper) pairing part. In the momentum space ($\bs{k}$-space), the BdG Hamiltonian of the $s$-wave superconductor is expressed as:
\be
\begin{split}
&H_{\rm BdG} = \sum_{\bs{k}} \Psi_{\bs{k}}^\dagger H_{{\rm BdG},\bs{k}}(\Delta) \Psi_{\bs{k}},\\{\rm where}~ &H_{{\rm BdG},\bs{k}}(\Delta) = \epsilon(|\bs{k}|)\,\tau_z \otimes \mathbbm{1}_{2 \times 2} - \left({\rm Re}\, \Delta\right)\,\tau_y \otimes \sigma_y - \left({\rm Im}\, \Delta\right)\,\tau_x \otimes \sigma_y.
\end{split}
\ee 
Here, the spinor $\Psi_{\bs{k}} = \left(c_{\bs{k}\uparrow}\,\, c_{\bs{k}\downarrow}\,\, c_{-\bs{k}\uparrow}^\dagger \,\, c_{\bs{k}\downarrow}^\dagger\right)^{\rm T}$ and the Pauli matrices $\tau_\alpha$ and $\sigma_\alpha$ are defined respectively in the particle-hole degrees of freedom and the spin degrees of freedom. The momentum dependent function $\epsilon(|\bs{k}|) = \frac{|\bs{k}|^2}{2m} - \mu$, where $\mu$ is the chemical potential. The last two terms in the above Hamiltonian is coming from the momentum independent $s$-wave pairing. The complex number $\Delta$ describes the physics of the pairing of two electrons to form a Cooper pair or the pairing of two holes to represent the breaking apart of the Cooper pair to its constituents. At the mean-field approximation limit, this represents the order parameter of the superconductor. The BdG Hamiltonian $H_{{\rm BdG},\bs{k}}(\Delta)$ are $4 \times 4$ matrices, but its energy spectrum has two doubly degenerate bands. Importantly, the Hamiltonian of this system is formed by the linear combination of the operators form a closed Lie algebra [a sub-algebra of $SU(2) \otimes SU(2)$].

\subsubsection{Unconventional ($p$-wave) superconductors}
 
We consider a simplest model of topological superconductor which is described by the mean-field BdG Hamiltonian of spin-less fermions. The spin-less fermions can be viewed as fully spin-polarized that happens due to the time-reversal symmetry (TRS) breaking in the system. The TRS breaking is physically possible in presence of magnetic field. We now discuss both 1D and 2D models of the $p$-wave superconductors.\\

\noindent \underline{1D case: $p$-wave wire}\\

Unlike the $s$-wave superconductor, for the spinless case, the $p$-wave pairing part of the Hamiltonian is dependent on the momentum. Thus the form of the BdG Hamiltonian in the $k$-space becomes
\be
\begin{split}
&H_{\rm BdG} = \frac{1}{2} \sum_k \Psi_k^\dagger H_{{\rm BdG},\bs{k}}(\Delta) \Psi_k\\
{\rm where}~ &H_{{\rm BdG},\bs{k}}(\Delta) = \epsilon(k)\, \tau_z + k \left({\rm Re}\, \Delta\right) \tau_x - k \left({\rm Im}\, \Delta\right) \tau_y.
\end{split}
\ee
Here $\Psi_k = \left(c_k~c_k^\dagger\right)^{\rm T}$ and like earlier $\epsilon(k) = \frac{k^2}{2m} - \mu$. This Hamiltonian has two energy bands $E_\pm = \pm \sqrt{\epsilon(k)^2 + k^2 |\Delta|^2}$. This dispersion relation can also be achieved starting from the 1D Kitaev chain \cite{Kitaev2001}. The Kitaev chain is the simplest model which illustrates the topological superconductor. The Hamiltonian of the Kitaev chain is given as
\be
H = \sum_j \left[\mu \left(2 c_j^\dagger c_j -1\right) - t \left(c_j^\dagger c_{j+1} + c_{j+1}^\dagger c_j\right) - |\Delta| \left(c_{j+1}^\dagger c_j^\dagger + c_j c_{j+1}\right)\right],
\ee
where $t$ is the hopping probability to the nearest neighbor sites, $\mu$ is the chemical potential, and the parameter $\Delta$ is chosen as real for simplicity. The Fourier transformation of this Hamiltonian in the $k$-space becomes
\be
H_{\rm Kitaev} = \sum_k \Psi_k^\dagger H_{{\rm Kitaev}, k} \left(|\Delta|\right) \Psi_k, ~~{\rm where}~H_{{\rm Kitaev}, k} \left(|\Delta|\right) =  \left[\mu-t\cos k\right] \tau_z - (|\Delta| \sin k)\, \tau_y,
\ee
where $\tau_\alpha$ are the pseudo-spin operators defined in the particle-hole degrees of freedom. Around $k \sim 0$, the Kitaev Hamiltonian gives equivalent band dispersion relation as that of the BdG Hamiltonian for the real $\Delta$. \\

\noindent \underline{1D case: Anisotropic $XY$ model in an external transverse field}\\

The Hamiltonian of the anisotopic $XY$ model in an external transverse field is given as
\be
H_{XY} = -\sum_j \left[\mu \sigma_j^z - J_x \sigma_j^x \sigma_{j+1}^x - J_y \sigma_j^y \sigma_{j+1}^y\right].
\ee
This Hamiltonian can be mapped exactly onto the above Kitaev Hamiltonian by means of Jordan-Wigner transformation. The anisotropy parameters are related to the hopping probability $t$ and the pairing parameter $|\Delta|$ as $J_x = \frac{1}{2}\left(t+|\Delta|\right)$ and $J_y = \frac{1}{2}\left(t-|\Delta|\right)$. Therefore, the anisotropic $XY$-model can be represented in terms of the generators of the $SU(2)$ algebra by two consecutive transformations: first Jordan-Wigner transformation, which transforms the spin system to a spin-less ferimionic system (Kitaev model); and then a Fourier transformation will transform this system into the momentum space.  

Here, we have given couple of examples of interacting system which come under the purview of our Floquet engineering protocol.\\

\noindent \underline{2D case: Chiral $p$-wave superconductor}\\

We now move to the higher dimensional version of the unconventional superconductor, i.e., 2D chiral $p$-wave superconductor. This system is very interesting because its vortices exhibit anyon excitations \cite{Volovik1999,Read-Green,Ivanov}. The Hamiltonian of this system in the real space is \cite{Kitaev2001}
\be
\begin{split}
H =& \sum_{m, n} \left[ -t \left(c_{m+1, n}^\dagger \,c_{m, n} + {\rm H.c}\right) - t \left(c_{m, n+1}^\dagger \,c_{m, n} + {\rm H.c}\right) - (\mu-4t) c_{m, n}^\dagger \,c_{m, n}\right.\\ &\left.+ \left(\Delta c_{m+1, n}^\dagger \,c_{m, n}^\dagger + \Delta^* c_{m, n}^\dagger \,c_{m+1, n}\right) + \left(i\Delta c_{m, n+1}^\dagger \,c_{m, n}^\dagger - i\Delta^* c_{m, n}^\dagger \,c_{m, n+1}\right)\right].
\end{split}
\ee  
The fermion operators $c_{m,n} \left(c_{m,n}^\dagger\right)$ annihilate (create) spinless fermions at the $(n,m)$ site. The pairing amplitude (complex $\Delta$) is anisotropic and an additional phase $i=e^{i\pi/2}$ is introduced along the $y$-direction. Like the 1D case, here also the pairing is not onsite, therefore the pairing will be momentum dependent. As usual, we can go from the 2D lattice space to the momentum space by Fourier transformation. Then assuming $\Delta = |\Delta|\, e^{i\theta}$ and making a gauge transformation $\left(c_{\bs{k}}, c_{\bs{k}}^\dagger\right) \rightarrow \left(c_{\bs{k}}\,e^{i\theta/2}, c_{\bs{k}}^\dagger\,e^{-i\theta/2} \right)$, we write the Hamiltonian in the BdG form as (in $\bs{k}$-space)
\be
H_{BdG} = \frac{1}{2} \sum_k \Psi_{\bs{k}}^\dagger H_{{\rm BdG}, \bs{k}} \Psi_{\bs{k}},~~{\rm where}~H_{{\rm BdG}, \bs{k}} = [2-\mu-\cos (k_x) -\cos( k_y)]\,\tau_z - 2 |\Delta| \sin (k_x) \,\tau_y - 2 |\Delta| \sin (k_y) \,\tau_x.
\ee
This is an example of a 2D interacting system which at the mean-field level can be represented by the linear combination of the pseudo-spin operators which form a closed Lie algebra. Therefore, the Floquet engineering protocol presented in this letter is also applicable for this important class of interacting systems.


\section{Three bands case}

In order to represent any generic three bands tight-binding Hamiltonian, one needs the $3 \times 3$ identity matrix $\mathbbm{1}$ and {\it eight} linearly independent matrices. A natural choice for these is to consider the eight trace-less Hermitian Gell-Mann matrices used in the standard description of $SU(3)$ algebra \cite{Gell-Mann,Nee-Mann}. The Gell-Mann matrices are generalizations of the Pauli matrices for the $3 \times 3$ case. In the standard basis, the Gell-Mann matrices are of the form:
\be
\begin{split}
\lambda_1 &= \begin{pmatrix} {0} & {1} & {0}\\{1}&{0}&{0}\\{0}&{0}&{0}\end{pmatrix},~~
\lambda_2 = \begin{pmatrix} {0}&{-i}&{0}\\{i}&{0}&{0}\\{0}&{0}&{0}\end{pmatrix},~~
\lambda_3 = \begin{pmatrix} {1}&{0}&{0}\\{0}&{-1}&{0}\\{0}&{0}&{0}\end{pmatrix},~~
\lambda_4 = \begin{pmatrix} {0}&{0}&{1}\\{0}&{0}&{0}\\{1}&{0}&{0}\end{pmatrix},\\
\lambda_5 &= \begin{pmatrix} {0}&{0}&{-i}\\{0}&{0}&{0}\\{i}&{0}&{0}\end{pmatrix},~~
\lambda_6 = \begin{pmatrix} {0}&{0}&{0}\\{0}&{0}&{1}\\{0}&{1}&{0}\end{pmatrix},~~
\lambda_7 = \begin{pmatrix} {0}&{0}&{0}\\{0}&{0}&{-i}\\{0}&{i}&{0}\end{pmatrix},~~
\lambda_8 = \frac{1}{\sqrt{3}} \begin{pmatrix} {1}&{0}&{0}\\{0}&{1}&{0}\\{0}&{0}&{-2}\end{pmatrix}. 
\end{split}
\ee 
In terms of the above matrices, one can write the Hamiltonian of any three-bands models such as \cite{Mizoguchi}
\begin{eqnarray}
\mathcal{H}_{\bs{k}} &=& h_{{\bs k}1} \lambda_1 + h_{{\bs k}4} \lambda_4 + h_{{\bs k}6} \lambda_6 , \quad {\rm (Kagome~Lattice)}\\
\mathcal{H}_{\bs{k}} &=&  h_{{\bs k}4}^\prime \lambda_4 + h_{{\bs k}6}^\prime \lambda_6 \quad {(\rm Lieb~Lattice)}.
\end{eqnarray}
If one wants to study Floquet version of the Kagome or Lieb lattice under the Wei-Norman formalism, then one has to consider all the $\lambda$-matrices. Following the Wei-Norman ansatz, the micro-motion operator will take the form
\be
P_{\bs{k}}(t) = e^{-im_{\bs{k}0}(t)\mathbbm{1}} \prod_{\alpha=1}^8 e^{-i m_{\bs{k}\alpha}(t) \lambda_\alpha}.
\ee 
Alternatively, one can construct the micro-motion operator using the following representation \cite{ARPRau-SU3}:
\be
a_\pm = \frac{1}{2} (\lambda_6 \pm i \lambda_7),~b_\pm = \frac{1}{2} (\lambda_1 \pm i \lambda_2),~c_\pm = \frac{1}{2} (\lambda_4 \pm i \lambda_5),~a_3 = \frac{1}{2} (\sqrt{3}\lambda_8 - \lambda_3),~{\rm and}~c_3 = \frac{1}{2} (\sqrt{3}\lambda_8 + \lambda_3).
\ee
In principle, one can follow our Floquet engineering protocol to realize any three-bands model using one of the above (or any other) representations of the $SU(3)$ algebra. However, the matrices $\bs{\mathcal{M}_{k}}_1(t)$ and $\bs{\mathcal{M}_{k}}_2(t)$ which are crucial for designing the driving protocol will now be $8 \times 8$ matrices. The large dimension of these matrices makes the Floquet engineering protocol for the three-bands case complicated.    
 
Interestingly, a careful observation reveals that $\lambda_1, \lambda_2,$ and $\lambda_3$ can be represented in terms of the Pauli matrices as 
\be
\lambda_\alpha = \begin{pmatrix} \sigma_\alpha & \bs{0} \\ 0 & 0 \end{pmatrix},~~{\rm where}~~ \alpha = 1, 2, 3~{\rm or}~x, y, z ~~{\rm and}~~\bs{0} = \begin{pmatrix} 0 \\ 0 \end{pmatrix}.
\ee
These three matrices form an $SU(2)$ sub-algebra: $[\lambda_\alpha, \lambda_\beta] = i \epsilon_{\alpha\beta\gamma} \lambda_\gamma$. Therefore, if one wants to design a Floquet engineering protocol for a system having three energy bands whose $H^{\rm eff}_{\bs{k}}$ can be expressed as a linear combination of these three matrices and the $3 \times 3$ identity matrix, then the undriven Hamiltonian $H_{\bs{k}0}$ and the periodic driving $V_{\bs{k}}(t)$ can also be expanded as a linear combination of the same. Instead of this representation, one can also use another representation, which is equivalent to the $\pm Z$ representation of the main text, with
\be
b_\pm = \frac{1}{2} \begin{pmatrix} \sigma_x \pm i \sigma_y & \bs{0}\\ 0 & 0 \end{pmatrix} = \begin{pmatrix} S_\pm & \bs{0}\\ 0 & 0 \end{pmatrix} \equiv \Lambda_\pm ~~{\rm and}~~\frac{1}{2} \lambda_3 = \frac{1}{2}  \begin{pmatrix} \sigma_z  & \bs{0}\\ 0 & 0 \end{pmatrix} = \begin{pmatrix} S_z  & \bs{0}\\ 0 & 0 \end{pmatrix} \equiv \Lambda_z.
\ee
to express the Hamiltonian. Moreover, like $S_\pm^2 = 0$, we now have $\Lambda_\pm^2 = 0$. Following the Wei-Norman ansatz, we can write down the form of the micro-motion operator for this case as 
\be
P_{\bs{k}}(t) = e^{-im_{\bs{k}0}(t)\mathbbm{1}} e^{-im_{\bs{k}+}(t) \Lambda_+} e^{-im_{\bs{k}-}(t) \Lambda_-} e^{- i m_{\bs{k}z}(t) \Lambda_z}.
\ee 
Here again, the operators $\Lambda_\pm$ are not Hermitian, and consequently the second and the third term in the above expression are not unitary. The unitary property of the micro-motion operator $P_{\bs{k}}(t)$ once again gives the same relation as given in Eq. \eqref{eq:dep_var}. As a consequence, from the time-derivative of the above $P_{\bs{k}}(t)$, we construct $\bs{\mathcal{M}_{k}}_1(t)$ matrix for the $SU(2)$ sub-algebra of the $SU(3)$ algebra. This $\bs{\mathcal{M}_{k}}_1(t)$ will be exactly identical to the expression given in Eq. (16) of the main text. The form of the $\bs{\mathcal{M}_{k}}_2(t)$ matrix is determined by the desired Hamiltonian $H^{\rm eff}_{\bs{k}}$. 

We may consider one interesting case for this $SU(2)$ sub-algebra. Consider a desired $3$-bands Hamiltonian of the form
\be
H^{\rm eff}_{\bs{k}} = \eta_{\bs{k}x} \Lambda_x + \eta_{\bs{k}y} \Lambda_y + \eta_{\bs{k}z} \Lambda_z = \eta_{\bs{k}-} \Lambda_+ + \eta_{\bs{k}+} \Lambda_- + \eta_{\bs{k}z} \Lambda_z \equiv \bs{h}_{\bs{k}}^{\rm eff} \bs{\cdot} \bs{\Lambda},
\ee
where \[\Lambda_\alpha = \frac{1}{2} \lambda_\alpha,~\alpha = x,\,y,\,z ~{\rm and}~\eta_{\bs{k}\pm} = \frac{1}{2}(\eta_{\bs{k}x} \pm i \eta_{\bs{k}y})\] and in $\pm Z$ representation $\bs{h}_{\bs{k}}^{\rm eff} = (\eta_{\bs{k}-},\,\eta_{\bs{k}+},\, \eta_{\bs{k}z})$. For the above Hamiltonian, two bands will be dispersive \[E_{\bs{k}} =  \pm \sqrt{|\eta_{\bs{k}+}|^2 + \frac{\eta_{\bs{k}z}^2}{4}} = \pm \frac{1}{2} \sqrt{\eta_{\bs{k}x}^2 + \eta_{\bs{k}y}^2 + \eta_{\bs{k}z}^2} \equiv \pm \frac{1}{2} |\bs{\eta}_{\bs{k}}|\] and the third band will be a flat-band at $E_{\bs{k}} = 0$. If we add a term $\eta_0 \mathbbm{1}$ to the above Hamiltonian, then the flat-band will be at energy $E_{\bs{k}} = \eta_0$ and the dispersive bands will be $E_{\bs{k}} = \eta_0 \pm \frac{1}{2} |\bs{\eta}_{\bs{k}}|$.
 
We now design the driving protocol to achieve the desired/effective  Hamiltonian $H^{\rm eff}_{\bs{k}}$. For simplicity, we are assuming the case when $\eta_0 = 0$, that is the energy of the flat-band is {\it zero}. Therefore, for this case, we can assume that there is no initial static Hamiltonian. We only need a pure time time-dependent Hamiltonian for any momentum $\bs{k}$ as
\be
H_{\bs{k}}(t) = F_{\bs{k}x}(t)\, \Lambda_x + F_{\bs{k}y}(t)\, \Lambda_y + F_{\bs{k}z}(t)\, \Lambda_z = F_{\bs{k}-}(t)\, \Lambda_+ + F_{\bs{k}+}(t)\, \Lambda_- + F_{\bs{k}z}(t)\, \Lambda_z.
\ee 
The driving functions $\{F_{\bs{k}x}(t), \,F_{\bs{k}y}(t), \,F_{\bs{k}z}(t)\}$ and  $\{F_{\bs{k}\pm}(t), \,F_{\bs{k}z}(t)\}$, where $F_{\bs{k}\pm}(t) = \frac{1}{2} \left[F_{\bs{k}x}(t) \pm i F_{\bs{k}y}(t)\right]$, are all time-periodic functions with period $T$.

The next is to derive $\bs{\mathcal{M}_{k}}_1(t)$ and $\bs{\mathcal{M}_{k}}_2(t)$ matrices for this case. Since we are considering only $SU(2)$ sub-group of the $SU(3)$ group, here also we get the identical $\bs{\mathcal{M}_{k}}_1(t)$ and $\bs{\mathcal{M}_{k}}_2(t)$ matrices as given by Eq. (15) of the main text. Moreover, if you consider the same gauge, then we shall also get identical $\bs{\tilde{\mathcal{M}}_{k}}_1(t)$ and $\bs{\tilde{\mathcal{M}_{k}}}_2(t)$ matrices as obtained in the main text [just below Eq. (17)]. In order to respect the boundary condition, we also set $\mu_+(t) = a_+ \sin (\omega t) = \mu_+^*(t)$ and $\mu_z^{\rm R}(t) = p\omega t$ where $p$ is any integer. We then obtain the driving functions in the $XYZ$ representation, using the relations  $F_{\bs{k}x} = 2{\rm Re} [F_{\bs{k}-}(t)]$ and $F_{\bs{k}y} = -2{\rm Im} [F_{\bs{k}-}(t)]$, as:

\begin{eqnarray}
F_{\bs{k}x}(t) &=&  2 F_e(t) \left[a_+ \omega C_{\omega t} C_{\bs{k}} - a_+ p\omega C_{\omega t} S_{\bs{k}} + \eta_{\bs{k}x} C_{p\omega t} - \eta_{\bs{k}y} S_{p\omega t} + a_+^2 S_{\omega t}^2 \left\{C_{2\bs{k} + p\omega t} \eta_{\bs{k}x} - S_{2\bs{k}+p\omega t} \eta_{\bs{k}y}\right\} - a_+ S_{\omega t} C_{\bs{k}} \eta_{\bs{k}z}\right],\nonumber\\
F_{\bs{k}y}(t) &=& -2 F_e(t) \left[a_+\omega C_{\omega t} S_{\bs{k}} + a_+p\omega C_{\omega t} C_{\bs{k}} - C_{p\omega t} \eta_{\bs{k}y} - S_{p\omega t} \eta_{\bs{k}x} + a_+ S_{\omega t}^2 \left\{S_{2\bs{k}+p\omega t} \eta_{\bs{k}x} - C_{2\bs{k}+p\omega t} \eta_{\bs{k}y}\right\} + a_+ S_{\omega t} C_{\bs{k}} \eta_{\bs{k}z}\right], \nonumber\\
F_{\bs{k}z}(t) &=& F_e(t) \left[2a_+ S_{\omega t} \eta_{\bs{k}y} + \left(1-\frac{a_+^2}{2}\right) p\omega + \frac{1}{2} p\omega a_+^2 C_{2\omega t} + \left\{\left(1-\frac{a_+^2}{2}\right) + \frac{a_+^2}{2} C_{2\omega t} \right\} \eta_{\bs{k}z} \right],
\end{eqnarray}
where $F_e(t) = (1 + a_+^2 S_{\omega t}^2)^{-1},\,C_w = \cos (w),$ and $S_w = \sin (w)$.  Here again we set $a_+^2 = 2$ and $p=3$ to remove the static part from the driving. We have shown the driving functions for the one dimensional case $\bs{k} \rightarrow k$ with $\eta_{kz} = 0$ for all $k$ and $\eta_{kx} = - \eta_{ky} = 2 \cos (k) + \Delta$. Here we again consider two values of the frequency, $\omega = 4 + 2 \Delta = 8$ and $\omega = 2 \Delta = 4$.

\begin{figure}[h]
\includegraphics[width=\columnwidth]{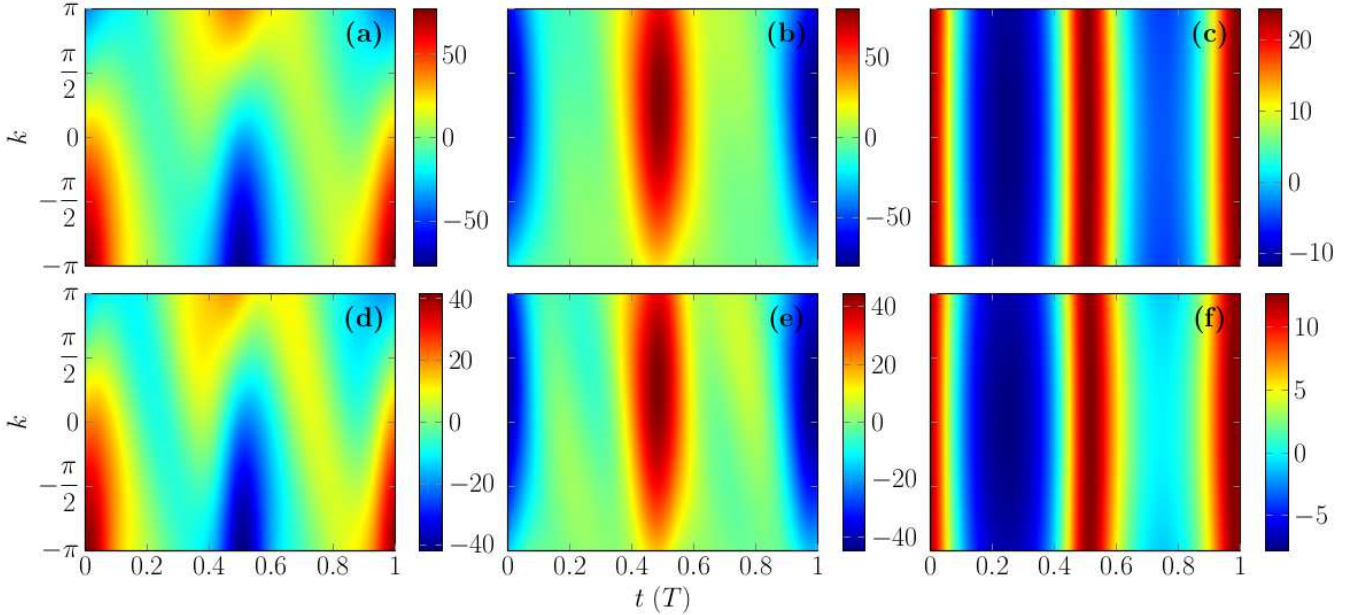}
\caption{Density plot of the driving functions $f_{kx}(t)$ (a,d), $f_{ky}(t)$ (b,e), $f_{kz}(t)$ (c,f) are plotted as a function of the time $t$ and the momentum $k$. Panels (a)-(c) represent $\omega = 4 + 2 \Delta = 8$ and panels (d)-(f) are for $\omega = 2 \Delta = 4$.} 
\label{fig:drive-3bands}
\end{figure}

In Fig. \ref{fig:drive-3bands}, we have shown the density plot of all the driving functions $f_{kx},\,f_{ky}$ and $f_{kz}$ on the plane of time $t$ and momentum $k$.

\end{appendix}
\end{widetext}

\bibliographystyle{apsrev4-2}

\begin{thebibliography}{54}
\bibitem{note}  For $p=1$, $f_{kz}$ develops a static contribution; whereas for $p=2$, $f_{kx}$ and $f_{ky}$ have static parts.
\end{thebibliography}
%
\end{document}